\documentclass[twocolumn,preprintnumbers,prl]{revtex4-1}
\usepackage{amsmath,amssymb,bm,mathrsfs}
\usepackage{hyperref}
\usepackage{url}
\usepackage{breakurl}
\usepackage{graphicx}

\newcommand{\beq}{\begin{equation}}
\newcommand{\eeq}{\end{equation}}
\newcommand{\beqn}{\begin{eqnarray}}
\newcommand{\eeqn}{\end{eqnarray}}

\def\rq{R_{\rm QCD}}
\def\rsq{R_{\rm SQCD}}
\def\os{O(\alpha_s}
\def\as{\alpha_s}

\newcommand{\gsim}{\lower.7ex\hbox{$\;\stackrel{\textstyle>}{\sim}\;$}}
\newcommand{\lsim}{\lower.7ex\hbox{$\;\stackrel{\textstyle<}{\sim}\;$}}

\begin{document}

\preprint{FTPI-MINN-14/41, UMN-TH-3412/14 }

\title{ Exact Adler Function in Supersymmetric QCD}
\author{M. Shifman}
 \affiliation{William I. Fine Theoretical Physics Institute, University of Minnesota,
Minneapolis, MN 55455, USA}

\author{K. Stepanyantz}%
\affiliation{M.V.Lomonosov Moscow State University, Faculty of Physics,
Leninskie Gory, Moscow 119991, Russia}


\begin{abstract}
The Adler function $D$ is found {\em exactly} in supersymmetric QCD.  Our exact formula
    relates $D(Q^2)$ to the anomalous dimension of the matter
    superfields $\gamma (\alpha_s(Q^2))$.   {\em En rout} we prove
     another theorem: the absence of the so-called singlet contribution to $D$. While such singlet
    contributions are present in individual supergraphs, they cancel in the sum.

\begin{description}

\item[PACS numbers]
11.15.-q, 11.30.Pb, 12.60.Jv, 12.38.-t
\end{description}
\end{abstract}

\maketitle

%\tableofcontents

\section{\label{sec:level1}Formulation of the problem and results}

The celebrated ratio $$R=\sigma (e^+e^-\to{\rm hadrons})/\sigma (e^+e^-\to \mu^+\mu^-)$$  plays a special role in QCD-based
phenomenology. For instance, it can be used for a precise determination of the gauge coupling $\alpha_s$
from accurate data on $e^+e^-\to$ hadrons in an appropriate energy range. It is also one of the key objects in various theoretical analyses in QCD, both in perturbation theory and beyond. In  perturbation theory
the ratio $R$ is defined as a normalized cross section
$$\sigma (e^+e^-\to \mbox{quarks + gluons})\,. $$
It is directly reducible to the imaginary part of the photon
polarization operator $\Pi$ (see (\ref{4})), \beq R_{\rm QCD} =
12\pi\,\mbox{Im}\, \Pi_{\rm QCD}\,. \eeq Alternatively, one can
define $\rq$ through a certain analytic continuation (see e.g.
\cite{ks}) of the Adler function \cite{Adler:1974gd}, \beq D (Q^2)
\equiv -  {12\pi^2}  \left(Q^2\, d/dQ^2 \right)\Pi (Q^2) \,,
\label{d2} \eeq In QCD the Adler function and the ratio $R$ are
calculated \cite{1} up to $\os^4)$. Supersymmetric QCD (SQCD) is
only a cousin of QCD since there is still no indication to the
existence of supersymmetry in our world. Nevertheless, SQCD is
known to be a unique theoretical laboratory in many aspects of
gauge dynamics. The $\os)$ correction to $R$ in SQCD was
calculated in \cite {2}.

In this paper we will
derive an exact relation between $D_{\rm SQCD}$ and the anomalous
dimension $\gamma$ of the matter superfield(s), valid to all
orders in $\alpha_s$,
 \beq D (Q^2)= \frac{3}{2}  N \sum_f
q_f^2\left[  1 -   \gamma \left(\alpha_s (Q^2) \right)\right]\,,
\label{m5} \eeq where $f$ is the flavor index, and $q_f$ is the
corresponding electric charge (in units of $e$). Equation
(\ref{m5}) assumes that all matter fields are in the fundamental
representation of $SU(N)$, although their electric charges can be
different. In calculating  $ \gamma \left(\alpha_s (Q^2) \right)$
one should remember that  $\alpha_s (Q^2) $ runs according to the
Novikov-Vainshtein-Shifman-Zakharov (NSVZ) $\beta$ function
\cite{nsvz,SV}. Our derivation of Eq. (\ref{m5}) refers to the
renormalization group (RG) functions defined in terms of the bare
coupling constant and uses the higher covariant
derivative regularization \cite{Slavnov_HD}.

From the practical side our result means, among other things, that
for this renormalization prescription $\os^n)$ calculation of the
Adler function $D(Q^2)$ in SQCD exactly reduces to a much simpler
$\os^{n-1})$ calculation of the anomalous dimension $\gamma$.

The photon polarization operator $\Pi (Q^2)$ is defined as
\beqn
\Pi_{\mu\nu} (q) &=&  i\int d^4 x \, e^{iqx} \left\langle T \left\{  j_\mu (x)\, j_\nu (0) \right\} \right\rangle
\nonumber\\
&\equiv& \left( q_\mu q_\nu - q^2 g_{\mu\nu} \right) \Pi (Q^2)\,,
 \label{4}
\eeqn where $Q^2 = - q^2$ and $j_\mu$ is the electric current. In
the case of QCD $j^\mu = \sum_f q_f \,\bar\psi_f \gamma^\mu
\psi^f$. In the supersymmetric case it is also necessary to take
into account the quarks' superpartners, squarks.

\begin{figure}
\includegraphics[scale=0.8]{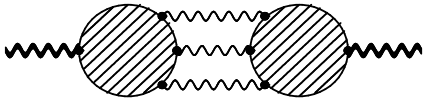}
\caption{\label{sc}An example of the singlet contribution to $D$. The shaded circles represent matter loops with all possible
$\alpha_s$ corrections. Thin wavy lines denote the gauge superfield $V$ while thick wavy lines denote external photons.}
\end{figure}

$\Pi (Q^2)$ consists of two parts. The so-called singlet part of
$\Pi$ is determined by graphs with at least two matter loops, with
photons attached to different loops, and is proportional to
$\left(\sum_f q_f\right)^2$, see Fig.~\ref{sc}. In the nonsinglet
part both external photon lines are attached to one and the same
matter loop; therefore, the nonsinglet part  is proportional to
$\left(\sum_f q_f^2\right)$. Correspondingly,
\begin{equation}
\label{Singlet}
D(\alpha_s) = \sum\limits_f q_f^2\, D_1(\alpha_s) +
\Big(\sum\limits_f\limits q_f\Big)^2 D_2(\alpha_s)\,.
\end{equation}
In deriving Eq. (\ref{m5}), {\em en rout} we explicitly establish
the following theorem:  it turns out that the singlet
contribution, symbolically depicted in Fig. 1, vanishes,
$D_2\equiv 0$,  once all relevant supergraphs are summed.  Only
the nonsinglet part $D_1$ survives. (This theorem was implicit in
\cite{svz}.) Hereafter, we will focus exclusively on $D_1$ keeping
in mind that  $D_2\equiv 0$.

A general derivation of the formula (\ref{m5}), relating $D$ and
$\gamma$, which is conceptually similar to the NSVZ $\beta$
function \cite{nsvz}, is based on an examination of a certain
``hybrid" $\beta$ function in SQCD, to be explained below, and
parallels the Shifman-Vainshtein nonrenormalization theorem
\cite{SV}.

When one deals with higher order corrections one must be careful
since higher-order terms in the perturbative expansion are
regularization and scheme dependent, generally speaking. Equation
(\ref{m5}) implies supersymmetric regularization as well as
renormalization scheme necessary for the NSVZ $\beta$ function.
The both elements were worked out in detail in \cite{3}. The
appropriate regularization is based on the higher derivative
method \cite{Slavnov_HD} supplemented by the Pauli-Villars
regularization for one-loop divergent (sub)diagrams
\cite{Slavnov_PV}.

Our general analysis of the Adler function is followed by a direct
supergraph calculation and comparison of $D(Q^2)$ and $\gamma$
which runs in parallel to that in  \cite{Stepanyantz:2011jy}. This
highly nontrivial calculation  fully confirms Eq. (\ref{m5}) -- a
considerable technical achievement in itself.

\section{\label{sec2}The model}

We will consider ${\mathcal N}=1$ SQCD with $N$ colors and $N_f$
flavors, assuming $N_f > N+1$. The latter condition is needed in
order to avoid nonperturbative quantum deformations of the moduli
space \cite{qd}. This will allow us to work at the origin of the
moduli space.

Each flavor is described by two chiral superfields $\Phi^i$ and
$\widetilde \Phi_i$ ($i$ is the color index) in the fundamental (antifundamental)
representations of $SU(N)$, respectively.
\begin{eqnarray}
&& S = S_{\mbox{\scriptsize gauge}} + S_{\mbox{\scriptsize
matter}}  = \frac{1}{2g_{0}^2}\mbox{Re} \, \mbox{tr} \int d^4x\,
d^2\theta\,W^2\nonumber\\
&& + \frac{1}{4 e_0^2} \mbox{Re} \int d^4x\, d^2\theta\,
\mbox{\boldmath$W$}^2  + \sum\limits_{f=1}^{N_f} \frac{1}{4}
\int d^4x\, d^4\theta\,\Big(\Phi_f^+ \nonumber\\
&& \times  e^{2q_f \mbox{\scriptsize \boldmath$V$}+2V}\Phi_f +
\widetilde\Phi_f^+ e^{-2q_f \mbox{\scriptsize \boldmath$V$}-2V^t}
\widetilde\Phi_f\Big)\,.\vphantom{\frac{1}{2}} \label{8}
\end{eqnarray}
The gauge sector consists of the dynamical $SU(N)$ part and an
auxiliary $U(1)$ part. The $U(1)$  gauge superfield
$\mbox{\boldmath$V$}$ (containing the photon field) is treated as
an external field and is present only in the external lines, as in
Fig.~\ref{sc}. The $SU(N)$ and $U(1)$ gauge couplings are denoted
by $g$ and $e$, respectively; the subscript 0 marks their bare
(unrenormalized) values, i.e. the values at the ultraviolet
cut-off. The $U(1)$ field strength tensor corresponding to
$\mbox{\boldmath$V$}$   is $\mbox{\boldmath$W$}$,
\begin{equation}
\mbox{\boldmath$W$}_a = \frac{1}{4} \bar D^2 D_a
\mbox{\boldmath$V$}\,,\qquad W_a \equiv \frac{1}{8} \bar{D}^2 (e^{-2V} D_a e^{2V})\,.
\end{equation}
$\mbox{\boldmath$V$}$ is coupled to (s)quarks in the standard way, see the last line in (\ref{8}).  Our notation is similar to that in \cite{BL}. Moreover,
\beqn
&& \int\theta^2 d^2\theta =2\,,\qquad \int \theta^2\bar\theta^2 d^4\theta =4\,,
\nonumber\\[2mm]
&& V=   V^A t^A\,, \qquad {\rm tr} (t^A t^B) = \delta^{AB}/2\,.
\eeqn

We will discuss the $\beta$ function for $\alpha = e^2/4\pi$,
ignoring all orders in the electromagnetic coupling higher than
the leading order, while all orders in $\alpha_s = g^2/4\pi$ will
be taken into account. This $\beta$ function (referred to above as
hybrid) is defined and parametrized as follows:
\begin{eqnarray}
&& \alpha_0^{-2}\beta =- \frac{d \left( \alpha
(M_0)^{-1}\right)}{d\log M_0}\nonumber\\
&&\qquad\quad \equiv \frac{1}{\pi} \left[b +
b_1\frac{\alpha_{0s}}{\pi}+
b_2\left(\frac{\alpha_{0s}}{\pi}\right)^2 + ...\right]. \label{10}
\end{eqnarray}
Here $M_0$ is the ultraviolet cut-off. In differentiating with
respect to $\log M_0$ we keep the renormalized couplings
$\alpha_{s}$ and the normalization point $\mu$ fixed. We will say
that in this case $\beta$ is defined in terms of the bare coupling
constant. Alternatively, one can keep $\alpha_{0s}(M_0)$ fixed and
differentiate over $\log \mu$. Then we obtain $\beta$ defined in
terms of the renormalized coupling constants. Generally speaking,
these are two distinct schemes. The difference between these
definitions are discussed in \cite{3} in detail. Following
\cite{SV,nsvz}, we will use the former procedure.

In the leading order in $\alpha_s$
\beq
b=\left\{\begin{array}{l}
\frac{2N}{3}\,,\quad \mbox{one Dirac spinor}\\
\frac{N}{6} \,, \quad\ \mbox{one complex scalar}\\
N\,,\quad\,\,  \mbox{one supersymmetric flavor}\end{array}
\right.\,\,\,. \label{coe} \eeq Our task is to determine
$b_{1,2,...}$.

\section{\boldmath{$\beta$} versus \boldmath{$D$} and comments on derivation}

The  $\beta$ function in (\ref{10}) is obtained in a conventional
way starting from the two-point Green function of the superfield
${\mbox{\boldmath$V$}}$. Due to the $U(1)$ background gauge
invariance it is transversal,
\begin{eqnarray}
\label{Two_Point_Function} \Delta\Gamma^{(2)} &= &-
\frac{1}{16\pi} \int
\frac{d^4q}{(2\pi)^4}\,d^4\theta\,\mbox{\boldmath$V$}(\theta,-q)\,\partial^2\Pi_{1/2}
\mbox{\boldmath$V$}(\theta,q)\nonumber\\
&\times
&\Big(d^{-1}(\alpha_0,\alpha_{0s},M_0/Q)-\alpha_0^{-1}\Big),
\end{eqnarray}
where $\partial^2\Pi_{1/2} = -D^a \bar D^2 D_a/8$ denotes the
supersymmetric transversal projection operator.
% and $\Delta\Gamma =
%\Gamma- S$. 
In our notation
\begin{equation}
d^{-1}(\alpha_0,\alpha_{0s},M_0/Q)-\alpha_0^{-1} = 4\pi
\Pi(\alpha_{0s},M_0/Q).
\end{equation}
Differentiating this equation with respect to $\log M_0$ and
taking into account that $d^{-1}$ (as a function of the
renormalized coupling constants) is independent of $M_0$, we
obtain
\begin{equation}\label{Hybrid_Beta}
\alpha_0^{-2} \beta = 4\pi\frac{d}{d\log
M_0}\Pi\Big(\alpha_{0s}(\alpha_s,M_0/\mu),M_0/Q)\Big),
\end{equation}
where the limit $Q/M_0\to 0$ is assumed. Let us define the
function $\alpha_{0s}(Q)\equiv \alpha_{0s}(\alpha_s,M_0/Q)$ by
replacing $\mu \to Q$. Then Eq. (\ref{Hybrid_Beta}) can be
rewritten as a relation between the functions $\beta$ and $D$:
\begin{equation}\label{Beta_Vs_D}
\alpha_0^{-2} \beta = - 4\pi\frac{d}{d\log
Q}\Pi\Big(\alpha_{0s}(Q),M_0/Q)\Big) = \frac{2}{3\pi}
D(\alpha_{0s}).
\end{equation}
Hence, the hybrid $\beta$ and the Adler $D$ functions coincide
modulo the overall normalization. In this way we arrive at (\ref{m5}).

In the mid-1980s an exact relation for the NSVZ $\beta$ function
in SQCD was obtained \cite{nsvz}, \beq \alpha_{0s}^{-2}\beta  =\!-
\frac{1}{2\pi}\left[3N -\sum_f T(R_f)(1\!-\!\gamma_f)\right]
\left(1-\frac{N\alpha_{0s}}{2\pi} \right)^{-1}. \label{13} \eeq
Here $\gamma_f$'s are the anomalous dimensions of the matter
superfields in the representation $R_f$, \beq \gamma =
-\frac{d\log Z}{d\log M_0} \label{a8} \eeq and the coefficients
$T(R_f)$ are related to the quadratic Casimir operators $C(R_f)$,
\beq T(R) = C(R) \,\, \frac{\mbox{dim}(R)}{N^2-1}\,. \eeq A
similar formula in supersymmetric quantum electrodynamics (SQED)
with one electron was obtained in \cite{svz}, \beq
\alpha_0^{-2}\beta_{\rm SVZ} = \frac{1}{\pi} \left[1- \gamma
(\alpha_0) \right]. \label{14} \eeq Superficially, Eqs. (\ref{14})
and (\ref{m5}) have the same factors in the square brackets; in
fact, they are different: $\gamma (\alpha_{0s})$ is calculated in
SQCD while $\gamma (\alpha_0)$ in SQED.

In all the above cases the arguments of  \cite{SV,vs} tell us that only the first loop is ``normal," the
Wilsonean $\beta$ function is exhausted by one loop. In other words, the coefficient $b=N$ in (\ref{10}),  (\ref{coe})
is ``normal", while $b_{1,2,...}$ are due to the matter operator in the effective action. Naively it vanishes by virtue of the equations of motion, but the
Konishi anomaly \cite{konishi} converts it into the $U(1)$ gauge kinetic term,
and, therefore, all higher orders come from $\gamma$'s.

\section{Verifying at order \boldmath{$\os)$}}

One can easily verify the match of the coefficient $b_1$. To this
end  let us compare our prediction  (\ref{m5}) with the results of
\cite{2}
 \beq \rsq =\frac{3}{2} N \sum_f q_f^2
\left[ 1+  \frac{N^2-1}{2N} \frac{\alpha_s}{\pi} +
O(\alpha_s^2)\right] .
\eeq
Using the fact that \cite{vs} \beq
\gamma (\alpha_s) = -\frac{N^2-1}{2N}\frac{\alpha_s}{\pi} +\os^2),
\eeq and that in the first order in $\as$ the Adler function $D$
coincides with $R$ we reproduce (\ref{m5}) to order $O(\alpha_s)$.
The coefficient $b_1$ is scheme-independent while $b_{2,3,...}$
will depend on the renormalization scheme.

\section{Scheme dependence in higher orders}

In direct perturbative derivation of Eq. (\ref{m5}) one should
understand that all coefficients starting from $b_2$ are scheme
dependent. To obtain the NSVZ $\beta$ functions by using
dimensional reduction \cite{Siegel} one has to ensure a specially
tuned finite renormalization. It was verified that in three-
and four-loop orders such a renormalization exists
\cite{Jack:1996vg}, to be referred to as the NSVZ scheme. The NSVZ
scheme (in which the NSVZ relations are valid in all orders) was
explicitly constructed in \cite{3,kat} by using the higher
derivative regularization.

As was already mentioned, in this paper we also calculate
supergraphs using the higher derivatives method \cite{Slavnov_HD}
supplemented by the Pauli-Villars regularization for one-loop
divergent (sub)diagrams \cite{Slavnov_PV}. This procedure can be
formulated in a manifestly supersymmetric way
\cite{Supersymmetric_HD}. A possible version of the higher
derivative term is as follows. We introduce superfield $\Omega$
related to the gauge superfield $V$ as
\begin{equation}
e^{2V} \equiv e^{\Omega^+} e^\Omega.
\end{equation}
The superfield $\Omega$ allows one to construct the gauge
covariant supersymmetric derivatives
\begin{equation}
\nabla_a = e^{-\Omega^+} D_a e^{\Omega^+}; \qquad \nabla_{\dot a}
= e^{\Omega} \bar D_{\dot a} e^{-\Omega}.
\end{equation}
Using the superfield $\Omega$ and the above
covariant derivatives we construct  an appropriate
higher derivative term,
\begin{eqnarray}
S_\Lambda &=& \frac{1}{2g_{0}^2} \mbox{tr} \int d^4x\,d^2\theta\,
(e^\Omega W^a e^{-\Omega})\nonumber\\
&\times&  \Big[ R\Big(-\frac{\bar\nabla^2
\nabla^2}{16\Lambda^2}\Big)-1\Big] (e^\Omega W_a e^{-\Omega}),
\end{eqnarray}
where $\Lambda$ is a parameter with the dimension of mass, which
plays the role of the ultraviolet cutoff (later we set $\Lambda
=M_{\rm PV}=M_0$). The regulator $R$ should obey the constraints
$R(0)-1=0$ and $R(x)\to \infty$ for $x\to\infty$. For example, one
can choose $ R(x) = 1 + x^n\,. $ Needless to say,  it is necessary
to fix a gauge by adding the term $S_{\mbox{\scriptsize gf}}$ to
the action and introduce the corresponding ghosts with the action
$S_{\mbox{\scriptsize ghosts}}$. The one-loop divergences which
remain after introducing the higher derivative term are removed
by inserting the Pauli--Villars determinants into the generating
functional \cite{Slavnov_PV}.

The Adler function defined in terms of the bare coupling constant
\begin{equation}\label{Adler_Function_Bare}
D(\alpha_{0s}) \equiv - \frac{3\pi}{2} \frac{d}{d\log\Lambda}
\alpha_0^{-1}(\alpha,\alpha_s,\Lambda/\mu),
\end{equation}
can be obtained from the expression (\ref{Two_Point_Function}) by
making a substitution $\mbox{\boldmath$V$} \to \theta^4$:
\begin{equation}\label{Adler_Function_Final}
\frac{1}{3\pi^2}{\cal V}_4 \cdot D(\alpha_{0s}) =
\frac{d(\Delta\Gamma^{(2)})}{d\log\Lambda}\Big|_{\mbox{\scriptsize
\boldmath$V$}=\theta^4},
\end{equation}
where ${\cal V}_4\to \infty$ is the space-time volume. (Certainly,
it should be properly regularized, see \cite{Stepanyantz:2014ima}
for details.)

By definition, the function (\ref{Adler_Function_Bare}) is
scheme-independent for a fixed regularization \cite{3,kat}. Here
we argue that it is related to the anomalous dimension (\ref{a8})
(where $M_0$ should be replaced by $\Lambda$), which is also
defined in terms of the bare coupling constant. The anomalous
dimension defined by Eq. (\ref{a8}) also does not depend on the
subtraction scheme for a fixed regularization.

Thus, we showed that, if the higher
derivative regularization is used, the functions $D$ in
(\ref{Adler_Function_Bare}) and $\gamma$
are related as
\begin{equation}\label{Exact_Adler_Function}
D(\alpha_{0s}) = \frac{3}{2} N \sum\limits_f q_f^2
\Big[1-\gamma(\alpha_{0s})\Big]
\end{equation}
in all orders independently of the subtraction scheme. This
statement is an analog of a similar statement proved for the
$\beta$-function of ${\cal N}=1$ SQED in \cite{Stepanyantz:2011jy}
and of ${\cal N}=1$ SQED with $N_f$ flavors in
\cite{Stepanyantz:2014ima}.

The scheme dependent RG functions are defined
in terms of the renormalized coupling constant. In this case the
derivatives with respect to $\log \mu$ are calculated at fixed
values of the {\cal bare} coupling constant. Then the exact
expression for the $D$ function is valid only in a certain
subtraction scheme which can be constructed by imposing
boundary conditions similar to ones considered in
\cite{3,Kataev:2014gxa}.

\section{Summation of supergraphs}

To prove Eq. (\ref{Exact_Adler_Function}) we note that momentum
integrals giving the function $D$ are integrals of double total
derivatives if the higher derivative method is used for
regularization of supersymmetric theories. This implies that they
have the same structure as integrals giving the NSVZ $\beta$ functions
in supersymmetric theories which was first noted in
\cite{Total_Derivatives} and subsequently confirmed by other
calculations \cite{Total_Derivatives_Calculations}. Hence,
one of the momentum integrals can be calculated analytically and
the function $D$ in the $n$-th loop can be written as an integral
over $(n-1)$ loop momenta. This integral does not vanishes due to
singularities of the integrand, which appear due to the identity
\begin{equation}
\left[\frac{\partial}{\partial Q_\mu}, \frac{Q^\mu}{Q^4}\right] = 2\pi^2
\delta^4(Q),
\end{equation}
where $Q^\mu$ denotes the Euclidian momentum. The sum of the
singularities gives the term with the anomalous dimension in the exact
expression for the Adler function. Details of our calculation will
be given elsewhere \cite{big}. Here we outline only
main stages.

We will use the notation
\begin{eqnarray}
&& * \equiv \frac{1}{1-(e^{2V}-1)\bar D^2
D^2/16\partial^2},\nonumber\\
&& \widetilde
* = \frac{1}{1-(e^{-2V}-1)\bar D^2 D^2/16\partial^2}.
\end{eqnarray}
These expressions encode sequences of vertices and propagators on
the matter line (for $\Phi$ and $\widetilde\Phi$, respectively).
Then the singlet contribution to the Adler function (after the
substitution $\bm{V}\to\theta^4$) is proportional to
\begin{eqnarray}\label{Singlet}
&& \frac{d}{d\log\Lambda} \Big\langle \Big[i \sum\limits_f q_f\,
\mbox{Tr} \Big(\bar\theta^c (\gamma^\mu)_c{}^d \theta_d [x_\mu,
\log(*) - \log(\widetilde
*)]\Big)\nonumber\\
&& + (PV) \Big]^2\Big\rangle = 0,
\end{eqnarray}
where $(PV)$ denotes the contribution of the Pauli--Villars
superfields. The commutator with $x_\mu$ corresponds to the
integral over the total derivative in the momentum space, which
vanishes because the integrand does not contain singularities. As
a consequence, the singlet contribution is given by integrals of
total derivatives and vanishes. (The Pauli--Villars contribution
has a similar structure and also vanishes for the same reason.)

The remaining contribution is proportional to
\begin{eqnarray}\label{NonSinglet}
&& i\frac{d}{d\log\Lambda} \sum\limits_f q_f^2
\mbox{Tr}\Big\langle\theta^4 \Big[x_\mu, \Big[x^\mu,\log(*)
+\log(\widetilde *) \Big] \Big]\Big\rangle
+(PV)\nonumber\\
&& - \mbox{terms with $\delta$-functions}.
\end{eqnarray}
Again,  this is an integral of a total derivative. However, it
does not vanish due to  singularities of the integrand. The
contribution of these singularities can be found repeating the
calculations made in \cite{Stepanyantz:2011jy}. It turns out that
it is proportional to the anomalous dimension of the matter
superfields and gives the second term in Eq.
(\ref{Exact_Adler_Function}). Details of this proof will be
published elsewhere \cite{big}.

%%%%%%%%%%%%%%%%%%%%%%%%%%%%%%%%%%%%%%%%%%%%%%%%%%%%%%%%%%%%
\vspace{1mm}
%\section*{Acknowledgments}
\begin{acknowledgments}
We are grateful to K. Chetyrkin and A. Kataev for useful
discussions.
The work of MS is supported in part by DOE grant DE-SC0011842. The
work of KS is supported by RFBR grant 14-01-00695.
\end{acknowledgments}


\begin{thebibliography}{99}

\bibitem{ks}
A.~A.~Pivovarov, Nuovo Cim.\ A {\bf 105}, 813 (1992); A.~L.~Kataev
and V.~V.~Starshenko, Mod.\ Phys.\ Lett.\ A {\bf 10}, 235  (1995);
A.~V.~Radyushkin, JINR Rapid Commun.\  {\bf 78}, 96 (1996).

\bibitem{Adler:1974gd}
S.~L.~Adler, Phys.\ Rev.\ D {\bf 10}, 3714 (1974).

\bibitem{1}
P.~A.~Baikov, K.~G.~Chetyrkin and J.~H.~K\"uhn, Phys.\ Rev.\
Lett.\  {\bf 101}, 012002 (2008); Phys.\ Rev.\ Lett.\  {\bf 104},
132004 (2010).

\bibitem{2}
A.~L.~Kataev and A.~A.~Pivovarov, JETP Lett.\  {\bf 38}, 369
(1983).

\bibitem{nsvz}
V.~A.~Novikov, M.~A.~Shifman, A.~I.~Vainshtein and V.~I.~Zakharov,
Nucl.\ Phys.\ B {\bf 229}, 381 (1983); Phys.\ Lett.\ B {\bf 166},
329 (1986).

\bibitem{SV}
M.~A.~Shifman and A.~I.~Vainshtein, Nucl.\ Phys.\ B {\bf 277}, 456
(1986).

\bibitem{Slavnov_HD}
A.~A.~Slavnov, Nucl.\ Phys.\ B {\bf 31}, 301 (1971); Theor. Math.
Phys. {\bf 13}, 1064 (1972).

\bibitem{svz}
M.~A.~Shifman, A.~I.~Vainshtein and V.~I.~Zakharov, Phys.\ Lett.\
B {\bf 166}, 334 (1986).

\bibitem{3}
A.~L.~Kataev and K.~V.~Stepanyantz, Nucl.\ Phys.\ B {\bf 875}, 459
(2013).

\bibitem{Slavnov_PV}
A.~A.~Slavnov, Theor.\ Math.\ Phys.\ {\bf 33} (1977) 977;
L.~D.~Faddeev and A.~A.~Slavnov, {\sl Gauge Fields: An
Introduction To Quantum Theory}, Second Edition (Westview Press,
1993).

\bibitem{Stepanyantz:2011jy}
K.~V.~Stepanyantz, Nucl.\ Phys.\ B {\bf 852}, 71 (2011).

\bibitem{qd}
N.~Seiberg, Phys.\ Rev.\ D {\bf 49}, 6857 (1994); Nucl.\ Phys.\ B
{\bf 435}, 129 (1995); K.~A.~Intriligator and N.~Seiberg, Nucl.\
Phys.\ Proc.\ Suppl.\ {\bf 45BC}, 1 (1996).

\bibitem{BL}
D. Bailin and A. Love, {\sl Supersymmetric Gauge Field Theory and
String Theory}, (IOP Publishing, Bristol, 1994).

\bibitem{vs}
M.~A.~Shifman and A.~I.~Vainshtein, Sov.\ J.\ Nucl.\ Phys.\  {\bf
44}, 321 (1986).

\bibitem{konishi}
K.~Konishi, Phys.\ Lett.\ B {\bf 135}, 439 (1984).

\bibitem{Siegel}
W.~Siegel, Phys.\ Lett.\ B {\bf 84}, 193 (1979).

\bibitem{Jack:1996vg}
I.~Jack, D.~R.~T.~Jones and C.~G.~North, Phys.\ Lett.\ B {\bf
386}, 138 (1996); Nucl.\ Phys.\ B {\bf 486}, 479 (1997); I.~Jack
and D.~R.~T.~Jones, Phys.\ Lett.\ B {\bf 415}, 383 (1997);
I.~Jack, D.~R.~T.~Jones and A.~Pickering, Phys.\ Lett.\ B {\bf
435}, 61 (1998).

\bibitem{kat}
A.~L.~Kataev and K.~V.~Stepanyantz, Phys.\ Lett.\ B {\bf 730}, 184
(2014).

\bibitem{Supersymmetric_HD}
V.~K.~Krivoshchekov, Theor.\ Math.\ Phys.\ {\bf 36}, 745 (1978);
P.~C.~West, Nucl.\ Phys.\ B {\bf 268}, (1986).

\bibitem{Stepanyantz:2014ima}
K.~V.~Stepanyantz, JHEP {\bf 1408}, 096,  (2014).

\bibitem{Kataev:2014gxa}
A.~L.~Kataev and K.~V.~Stepanyantz, arXiv:1405.7598 [hep-th].

\bibitem{Total_Derivatives}
A.~A.~Soloshenko and K.~V.~Stepanyantz, Theor.\ Math.\ Phys.\ {\bf
140}, 1264 (2004); A.~V.~Smilga and A.~Vainshtein, Nucl.\ Phys.\ B
{\bf 704}, 445 (2005).

\bibitem{Total_Derivatives_Calculations}
A.~B.~Pimenov, E.~S.~Shevtsova and K.~V.~Stepanyantz, Phys.\
Lett.\ B {\bf 686}, 293, (2010); K.~V.~Stepanyantz,
arXiv:1108.1491 [hep-th]; A.~E.~Kazantsev and K.~V.~Stepanyantz,
arXiv:1410.1133 [hep-th].

\bibitem{big}
M. Shifman and K. Stepanyantz, to appear.



\end{thebibliography}
\end{document}